\documentclass[%
mathleft,%
final,%
]{an}
\usepackage{graphicx}
\usepackage{times}

\begin{document}

\Pagespan{85}{}
\Yearpublication{2013}%
\Yearsubmission{2012}%
\Month{1}%
\Volume{334}%
\Issue{1/2}%
\DOI{10.1002/asna.201211783}%

\title{Infrared Parallaxes of Young Field Brown Dwarfs and Connections
  to Directly Imaged Gas-Giant Exoplanets}

\author{Michael C. Liu,\inst{1}\fnmsep\thanks{Corresponding author. 
  \email{mliu@ifa.hawii.edu}}
Trent J. Dupuy,\inst{2}
\and Katelyn N. Allers\inst{3}
}
\titlerunning{IR Parallaxes of Young Field Brown Dwarfs}
\authorrunning{M. Liu, T. Dupuy \& K. Allers}
\institute{
Institute for Astronomy, University of Hawaii, 2680
  Woodlawn Drive, Honolulu HI 96822 USA
\and
Harvard-Smithsonian Center for Astrophysics, 60 Garden
  Street, Cambridge, MA 02138 USA
\and
Department of Physics and Astronomy, Bucknell
  University, Lewisburg, PA 17837 USA
}

\received{September 12, 2012}
\accepted{January 9, 2013}


\abstract{%
  We have measured high-precision infrared parallaxes with the
  Canada-France-Hawaii Telescope for a large sample of candidate young
  ($\approx$10$-$100~Myr) and intermediate-age ($\approx$100$-$600~Myr)
  ultracool dwarfs, with spectral types ranging from M8 to T2.5. These
  objects are compelling benchmarks for substellar evolution and
  ultracool atmospheres at lower surface gravities (i.e., masses) than
  most of the field population.
  We find that the absolute magnitudes of our young sample can be
  systematically offset from ordinary (older) field dwarfs, with the
  young late-M objects being brighter and the young/dusty mid-L
  (L3--L6.5) objects being fainter, especially at $J$~band. Thus, we
  conclude the ``underluminosity'' of the young planetary-mass
  companions HR~8799b and 2MASS~J1207$-$39b compared to field dwarfs is
  also manifested in young free-floating brown dwarfs, though the effect
  is not as extreme. At the same time, some young objects over the full
  spectral type range of our sample are similar to field objects, and
  thus a simple correspondence between youth and magnitude offset
  relative to the field population appears to be lacking.
  Comparing the kinematics of our sample to nearby stellar associations
  and moving groups, we identify several new moving group members,
  including the first free-floating L~dwarf in the AB~Dor moving group,
  2MASS~J0355+11. Altogether, the effects of surface gravity (age) and
  dust content on the magnitudes and colors of substellar objects appear
  to be degenerate.}
\maketitle

\section{Introduction}

Direct detections of young exoplanets are now strengthening the link
between the exoplanet and brown dwarf populations, enriching our
understanding of both classes of objects. At the same time, recent
exoplanet discoveries display puzzling spectrophotometric properties,
with exceptionally red colors, peculiar near-IR spectra, and fainter
absolute magnitudes compared to field brown dwarfs (e.g.,
\cite{marois08-hr8799bcd}, \cite{bowler10-hr8799b},
\cite{2011ApJ...735L..39B}). Brown dwarfs have long been considered
valuable laboratories for discerning the physical properties of
gas-giant planets, which are much more difficult to study directly. And
yet the first examples of directly imaged planets, such as those around
the young A~star HR~8799 and the binary companion 2MASS~J1207$-$39b,
appear to be discrepant with field brown dwarfs. And thus we are faced
with the conundrum of how to integrate both classes of objects into a
common understanding of substellar evolution.

One promising approach is to identify robust free-floating analogs to
gas-giant planets in the solar neighborhood. The initial mass function
in the young star-forming clusters appear to go down to a several
Jupiter masses (e.g., \cite{2007A&A...470..903C},
\cite{2008MNRAS.383.1385L}), and thus such low mass objects should be
found in the field after departing their birth sites. At fixed effective
temperature, young ($\approx$10--100~Myr) field brown dwarfs will have
larger radii and lower masses than older field objects. The combination
of these two factors means a reduction in surface gravity by a factor of
$\approx$10. Some possible examples of young planet analogs are the rare
late-M and L-type field dwarfs with very red, dusty photospheres and/or
signs of low surface gravity in their optical and near-IR spectra. While
these objects have been studied spectroscopically, parallax measurements
have been lacking, representing a key information gap in characterizing
these objects.

%

\section{Observations}

Since 2007 we have been conducting a high-precision parallax program at
the 3.6-meter Canada-France-Hawaii Telescope (CFHT) using the facility
wide-field IR camera WIRCam (\cite{2004SPIE.5492..978P}). CFHT offers a
nearly ideal platform for parallaxes, given its combination of large
aperture, excellent seeing, and queue scheduling, though to our
knowledge it had not been used for parallaxes prior to our effort. As
described in Dupuy \& Liu (2012), our measurements are as good as have
ever been achieved in the near-IR, producing parallaxes with typical
uncertainties of 1.3~mas and as good as 0.7~mas, but for objects
$\approx$2--3~mags fainter than have been measured by previous work.

This combination of faint limiting infrared magnitudes \underline{and}
high precision
is relevant for studying young field brown dwarfs. Since such objects
are a small minority population,
their typical distances will be larger than ordinary (older) field
object. Furthermore, the stellar members of even the nearest young
moving groups can extended to distances of $\approx$60~pc (e.g.,
\cite{2008hsf2.book..757T}). This is quite far compared to previous
brown dwarf parallax programs. For instance, among $\ge$L4~dwarfs
(corresponding to the stellar/substellar boundary in the field), no
objects had high precision parallaxes ($\le3$\% uncertainties) beyond
13~pc prior to our CFHT effort.

Our ongoing CFHT program is monitoring candidate young field objects
with spectral types of M6 and later that have been identified from a
variety of sources, primarily candidate members of nearby young
($<$300~Myr) stellar associations and field objects whose optical and
near-IR spectral peculiarities are thought to arise from surface gravity
effects. Our sample contains three subsets:

\begin{enumerate}

\item {\em Low-gravity ultracool field objects:} We selected targets
  from SDSS or 2MASS-based searches for ultracool dwarfs in the solar
  neighborhood have been flagged as low gravity based on their optical
  and/or near-IR spectra (e.g.,
    \cite{2007AJ....133..439C}, 
    \cite{2008AJ....136.1290R}, 
    \cite{2008ApJ...689.1295K},
    \cite{2009ApJ...699..649S},
    \cite{2010ApJ...715..561A}).

\item {\em Stellar association members:} Our sample includes candidate
  members of the TW Hya Association (TWA; 8--10~Myr),
  Pleiades moving group (120~Myr),
  Ursa Major moving group (500~Myr),
  and Hyades cluster (625~Myr)
  These candidates have been identified in the literature based on their
  common kinematics with the stellar members of these groups.

\item {\em Extremely red L~dwarfs:} A small number of field L~dwarfs
  show extremely red near-IR colors and peculiar near-IR spectra. The
  archetype for this genre is 2MASS~J2244+20, which has an ordinary
  L6.5-type optical spectrum but a very unusual near-IR spectrum,
  distinguished by its much redder color, stronger CO, and more peaked
  $H$-band continuum shape compared to other field objects
  (\cite{2000AJ....120..447K}, \cite{2003ApJ...596..561M}). Kirkpatrick
  et~al.\ (2008) conclude this is young field object based on its
  near-IR SED, as opposed to an ordinary (high gravity) object with
  extreme clouds. Allers \& Liu (2012) scrutinize its near-IR spectrum
  and also conclude this object is young.

\end{enumerate}

Our methods for obtaining high-precision astrometry from CFHT/WIRCam
images are detailed in Dupuy \& Liu (2012). Our young sample has a time
baseline of $\approx$1.5--3.0 years with 7--12~observing epochs per
objects (e.g., Figure~1).
%
%
In all cases, the $\chi^2$ value of the fit (proper motion + parallax)
is commensurate with the number of degrees of freedom in each dataset,
validating the accuracy of our astrometric errors.
The median parallax uncertainty for the young sample presented here is
1.4~mas (5\% in the distances), with a median distance of 31~pc.

\begin{figure}
\hskip 0.1cm
\includegraphics[angle=0, width=8cm]{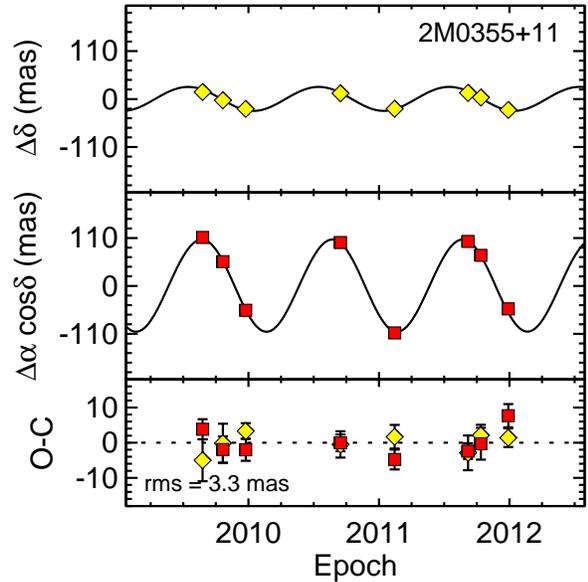}
\caption{One of our CFHT parallax measurements, for the young L5$\gamma$
  object 2MASS~J0355+11. The top and middle panels show relative
  astrometry in $\delta$ and $\alpha$, respectively, as a function of
  Julian year after subtracting the best-fit proper motion. (This is for
  display purposes only; our analysis fits for both the proper motion
  and parallax simultaneously.) The bottom panel shows the residuals
  after subtracting both the fitted parallax and proper motion, leaving
  a RMS scatter per epoch of 3.3~mas. After statistically accounting for
  the finite parallax of the background stars using a galactic
  population model, the resulting absolute parallax is
  $109.6\pm1.3$~mas, with $\chi^2=16.3$ (11 degrees of
  freedom). \label{fig:plx}}
\end{figure}


\section{Absolute Magnitudes}

Figure~2 shows the absolute magnitudes of our sample determined from
CFHT parallaxes, as a function of spectral type. Many of the young field
objects have $J$-band absolute magnitudes $M(J)$ that are displaced from
the locus of field dwarfs, but the displacement varies with spectral
type. The late-M young objects tend to be brighter in $M(J)$ than the
field objects, with the early-L dwarfs being comparable, and the mid-L
dwarfs being fainter. The faintest object in $J$-band is the primary
component of the young L3.5 binary SDSS~J2249+00A
(\cite{2010ApJ...715..561A}), which is about 1.5~magnitudes fainter than
field objects of comparable spectral type. Therefore, the fainter
$J$-band absolute magnitudes of the young planetary-mass companions
HR~8799b and 2MASS~J1207$-$39b compared to field dwarfs are also
manifested in young free-floating brown dwarfs.

However, perhaps a comparably interesting result is the large fraction
of our young targets which do {\em not} have significantly different
absolute magnitudes than the field sample, despite having spectroscopic
indications of low surface gravity. Almost none of our M9--L2 objects
are fainter than the field sequence within the measurement uncertainties
and the intrinsic scatter in the field population. In addition, within
the same spectral type, objects of different gravities appear to be
intermixed. A simple correspondence between spectroscopic diagnostics of
low surface gravity and $M(J)$ deviations is not present in our data.

\begin{figure}
\hskip 0.9cm
\includegraphics[angle=0, width=7.2cm]{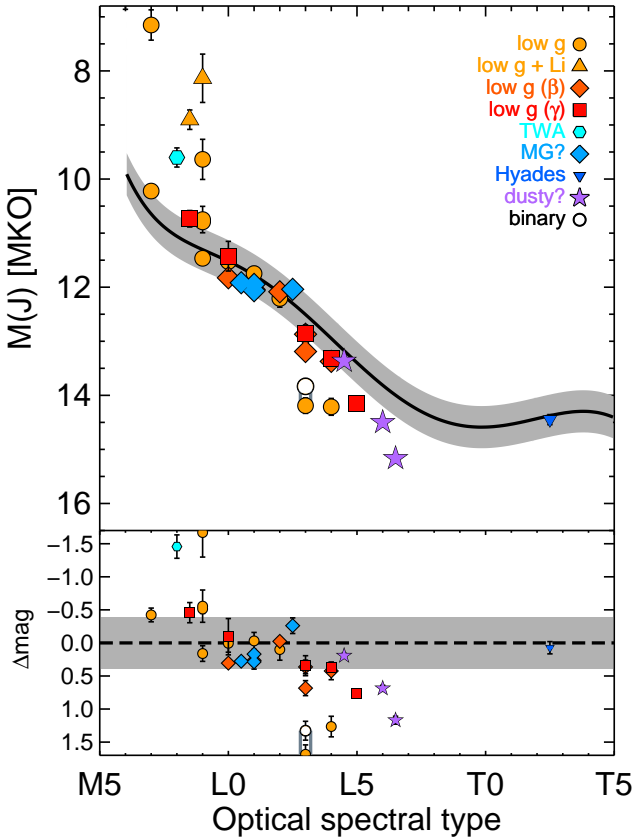}
\vskip 1.3cm
\caption{$J$-band absolute magnitudes of our young sample on the MKO
  system. The spectral types are all based on optical 
  data, except for the Hyades early-T~dwarf CFHT-Hy-20
  (\cite{2008A&A...481..661B}) which has a near-IR type. In the upper
  panel, the uncertainties in the absolute magnitudes are typically
  smaller than the plotting symbols. The thick black
  line shows the fit from Dupuy \& Liu (2012) for field
  ultracool dwarfs, and the light grey swath represents the 1$\sigma$
  scatter about the fit. The lower panel shows the difference of the
  data with respect to the polynomial fit. The integrated-light datum for
  the young binary SDSS~J2249+00AB (\cite{2010ApJ...715..561A}) is shown as an
  open circle and the resolved data for component~A as a filled
  colored circle, with grey vertical lines connecting the two circles.
  (We assume the optical spectral type of component~A is the same as the
  integrated-light type.)
  \label{fig:absmag}}
\end{figure}

Figure~3 shows the near-IR color-magnitude (CMD) diagram for our sample
compared to field dwarfs. Similar to the behavior as a function of
spectral type, the absolute magnitudes of our young objects reveal a
mixed picture. Many of them are displaced from the field sequence,
forming a brighter and/or redder locus. However, some objects are
intermingled with the field sequence in the near-IR CMD, despite their
low gravity optical spectra.

\begin{figure}
\includegraphics[angle=0, width=\linewidth]{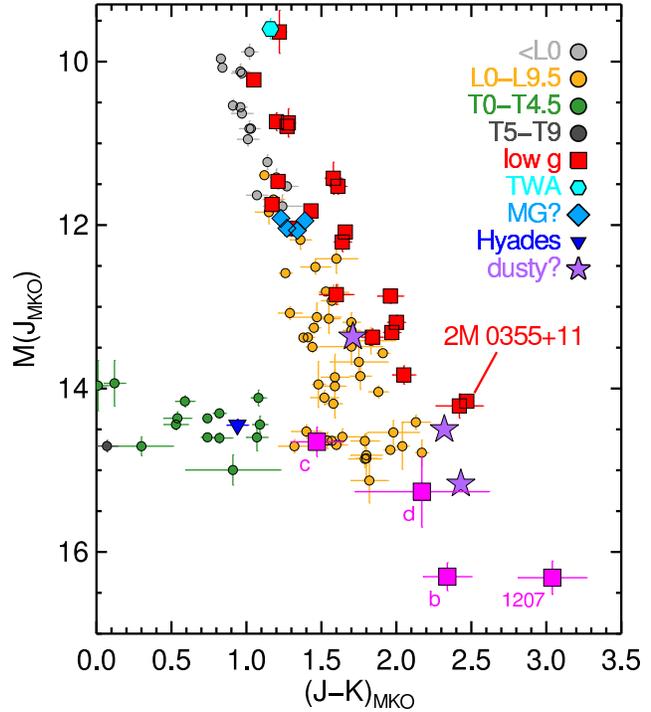}
\vskip 1ex
\caption{Infrared color-magnitude diagram on the MKO system showing our
  young sample compared to normal field dwarfs, with the latter from
  from Dupuy \& Liu (2012). The young substellar companions HD~8799bcd
  (\cite{marois08-hr8799bcd}) and 2MASS~J1207$-$39b
  (\cite{2005A&A...430.1027C}) are also plotted.  In this plot, low
  surface gravity objects of different classifications are all
  represented by the red square.  Candidate moving group members from the literature are labeled as ``MG?'' in the legend.
  \label{fig:cmd}}
\end{figure}

\section{New AB~Dor Member 2MASS~J0355+11}

By combining our parallaxes and proper motions with radial velocity data
in the literature, we can assess whether the 6-dimensional locations
(space velocity and position) of our targets are consistent with any
young moving groups or stellar associations. Such linkages would
establish the ages of the ultracool dwarfs, by adopting the ages
estimated for the stellar members, and thereby calibrate the
time dependence of gravity-dependent spectral features and delineate
empirical isochrones of substellar evolution. Also, such linkages
would add to the low-mass census of these groups, which is known to be
incomplete for optically faint members (e.g.,
\cite{2011ApJ...727....6S}).

We have identified several low-mass members of nearby young moving
groups, some previously flagged as candidates without using parallax
data and others as completely new linkages
(\cite{liu13-young-parallax}). One such object is the nearby L~dwarf
2MASS~J0355+11, one of the reddest L~dwarfs found from 2MASS by Reid
et~al.\ (2008) and optically classified as L5$\gamma$ by Cruz et~al.\
(2009). We measure a parallax of $109.6\pm1.3$~mas ($9.10\pm0.10$~pc),
making it the nearest known young brown dwarf.
%

Using the radial velocity from Blake et~al.\ (2010), we associate
2MASS~J0355+11 with the AB~Dor moving group, given its similar space
position to known members and the small $(U,V,W)$ difference
with the group (Figure 4). This suggests the $\gamma$ gravity
classification for mid-L dwarfs corresponds to an age of
$\approx$100~Myr, as determined for the AB~Dor moving group
(\cite{2008hsf2.book..757T}). This is older than the speculation of Cruz
et~al.\ (2009) that their lowest gravity objects ($\gamma$) have ages
closer to $\approx$10~Myr while the intermediate-gravity objects
($\beta$) are closer to $\approx$100~Myr.
In addition, this age estimate combined with the bolometric luminosity
of the object indicate a mass of $\approx$25~$M_{\rm Jup}$ based on
models by Chabrier et~al.\ (2000).

Contemporaneous with our Cool Stars~17 presentation, Faherty et~al.\ (2012)
presented a parallax of $134\pm12$~mas ($7.5^{+0.7}_{-0.6}$~pc) for
2MASS~J0355+11, based on infrared astrometry from the CTIO Blanco 4-m
Telescope. As a result, they find that 2MASS~J0355+11 is not associated
with any known young moving groups, thereby concluding that its age and
mass are indeterminate. The difference between their results and ours
arises from the parallaxes, with their CTIO and our CFHT measurements
differing by 2$\sigma$. While the statistical difference is relatively
modest, the $\approx$10$\times$ higher precision of our CFHT parallax
enables a more robust assessment of the kinematics.

Compared to field L~dwarfs, 2MASS~J0355+11 is redder than than the field
CMD locus (Figure 3) and fainter (0.7~mag) than other L5 dwarfs (Figure
2). However, it is still $\approx$2~magnitudes brighter than the
directly imaged planets. Indeed, there are other objects in our CFHT
sample that have even fainter absolute magnitudes and comparably red
colors, but their kinematics do not suggest they are young -- the
effects of dust and age appear to be degenerate in such data. Overall,
while some of the objects in our sample have atypical, or perhaps even
extreme, SEDs compared to most field objects, none of them coincide in
their colors and magnitudes with the exoplanets directly imaged to date.

\begin{figure}
\hskip 0.2in
\includegraphics[angle=0, width=7cm]{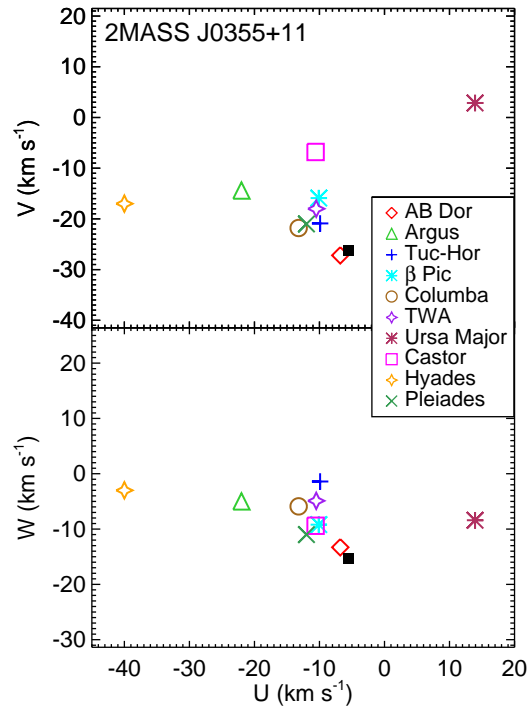}
\vskip -1ex
\caption{Space motion for 2MASS~J0355+11 (black square, with
  uncertainties smaller than the symbol size) compared to young moving
  groups and open clusters (colored symbols), based on our CFHT
  astrometry and the radial velocity from Blake et~al.\ (2010). For
  clarity, we have excluded the Columba, Octans, and $\eta$~Cha moving
  groups, as their locations are very similar to some of the groups
  plotted here but their sky location is inconsistent with our sample.
  Our measurement for 2MASS~J0355+11 agrees well with the AB~Dor moving
  group, with an offset of only $2.7\pm0.4$~km/s from the group's mean
  space motion. 
  Accounting for the internal velocity dispersion of the group, 
  we find a reduced $\chi^2$ value of 0.8 for 2MASS~J0355+11 being a
  member.
  \label{fig:uvw}}
\end{figure}

\acknowledgements This research was supported by NSF grants AST-0507833
and AST09-09222 (awarded to MCL) and NASA grant \hbox{{\#}HF-51271.01-A}
(awarded to TJD).


\begin{thebibliography}{22}
\expandafter\ifx\csname natexlab\endcsname\relax\def\natexlab#1{#1}\fi

\bibitem[{Allers \& {Liu} 2012}]{allers12-young-spectra}
Allers, K.~N., \& {Liu}, M.~C. 2012, \apj, submitted

\bibitem[{Allers {et~al.} 2010}]{2010ApJ...715..561A}
Allers, K.~N., {Liu}, M.~C., {Dupuy}, T.~J., \& {Cushing}, M.~C.
  2010, \apj, 715, 561

\bibitem[{{Barman} {et~al.} 2011}]{2011ApJ...735L..39B}
---. 2011, \apjl, 735, L39

\bibitem[{Blake {et~al.} 2010}]{2010ApJ...723..684B}
Blake, C.~H., {Charbonneau}, D., \& {White}, R.~J. 2010, \apj, 723, 684

\bibitem[{Bouvier {et~al.} 2008}]{2008A&A...481..661B}
Bouvier, J., {et~al.} 2008, A\&A, 481, 661

\bibitem[{Bowler {et~al.} 2010}]{bowler10-hr8799b}
Bowler, B., Liu, M., Dupuy, T., \& Cushing, M. 2010, \apj, 723, 850

\bibitem[{Caballero {et~al.} 2007}]{2007A&A...470..903C}
Caballero, J.~A. {et~al.} 2007, A\&A, 470, 903

\bibitem[{Chabrier {et~al.} 2000}]{2000ApJ...542..464C}
Chabrier, G., {Baraffe}, I., {Allard}, F., \& {Hauschildt}, P. 2000, \apj, 542,
  464

\bibitem[{Chauvin {et~al.} 2005}]{2005A&A...430.1027C}
Chauvin, G., {et~al.} 2005, A\&A, 430, 1027

\bibitem[{Cruz {et~al.} 2009}]{2009AJ....137.3345C}
Cruz, K.~L., {Kirkpatrick}, J.~D., \& {Burgasser}, A.~J. 2009, \aj, 137, 3345


\bibitem[{Cruz {et~al.} 2007}]{2007AJ....133..439C}
Cruz, K.~L., {et~al.} 2007, \aj, 133, 439

\bibitem[{Dupuy \& {Liu} 2012}]{dupuy11-parallax}
Dupuy, T.~J., \& {Liu}, M.~C. 2012, \apjs, 201, 19

\bibitem[{Faherty {et~al.} 2012}]{2012arXiv1206.5519F}
Faherty, J.~K., {Rice}, E.~L., {Cruz}, K.~L., {Mamajek}, E.~E., \&
  {N{\'u}{\~n}ez}, A. 2012, \aj, submitted

\bibitem[{Kirkpatrick {et~al.} 2000}]{2000AJ....120..447K}
Kirkpatrick, J.~D., {et~al.} 2000, \aj, 120, 447

\bibitem[{Kirkpatrick {et~al.} 2008}]{2008ApJ...689.1295K}
---. 2008, \apj, 689, 1295

\bibitem[{Liu {et~al.} 2013}]{liu13-young-parallax}
Liu, M.~C., Dupuy, T.~J., \& Allers, K.~N. {et~al.} 2013, ApJ, submitted

\bibitem[Lodieu {et~al.} 2008]{2008MNRAS.383.1385L}
Lodieu, N. {et~al.} 2008, \mnras, 383, 1385

\bibitem[{Marois {et~al.} 2008}]{marois08-hr8799bcd}
Marois, C., {Macintosh}, B., {Barman}, T., {Zuckerman}, B., {Song}, I.,
  {Patience}, J., {Lafreni{\`e}re}, D., \& {Doyon}, R. 2008, Science, 322, 1348

\bibitem[{McLean {et~al.} 2003}]{2003ApJ...596..561M}
McLean, I.~S., {McGovern}, M.~R., {Burgasser}, A.~J., {Kirkpatrick}, J.~D.,
  {Prato}, L., \& {Kim}, S.~S. 2003, \apj, 596, 561

\bibitem[{Puget {et~al.} 2004}]{2004SPIE.5492..978P} 
Puget, P., {et~al.} 2004, in SPIE Conference Series, Vol. 5492,
  Society of Photo-Optical Instrumentation Engineers (SPIE) Conference
  Series, ed. {A.~F.~M.~Moorwood \& M.~Iye}, 978--987

\bibitem[{Reid {et~al.} 2008}]{2008AJ....136.1290R}
Reid, I.~N., {et~al.} 2008, \aj, 136, 1290

\bibitem[{Shkolnik {et~al.} 2009}]{2009ApJ...699..649S}
Shkolnik, E., {Liu}, M.~C., \& {Reid}, I.~N. 2009, \apj, 699, 649

\bibitem[{Shkolnik {et~al.} 2011}]{2011ApJ...727....6S}
Shkolnik, E.~L., {Liu}, M.~C., {Reid}, I.~N., {Dupuy}, T., \& {Weinberger},
  A.~J. 2011, \apj, 727, 6

\bibitem[{Torres {et~al.} 2008}]{2008hsf2.book..757T}
Torres, C.~A.~O., {Quast}, G.~R., {Melo}, C.~H.~F., \& {Sterzik}, M.~F. 2008,
  {Young Nearby Loose Associations}, ed. {Reipurth, B.}, 757

\end{thebibliography}
\end{document}